\documentclass[12pt]{iopart}

\usepackage{cite,iopams}
\usepackage{latexsym}
\usepackage[dvips]{graphicx}
\usepackage{hyperref}

\newcommand{\ket}[1]{\left\vert#1\right\rangle}
\newcommand{\bra}[1]{\left\langle#1\right\vert}

\begin{document}

\title{Entanglement Entropy dynamics in Heisenberg chains}

\author{G De Chiara$^1$, S Montangero$^{1,2}$, P Calabrese$^3$ and R Fazio$^{1,4}$}

\address{$^1$ NEST-CNR-INFM \& Scuola Normale Superiore, Piazza dei
	Cavalieri 7, I-56126 Pisa, Italy}
\address{$^2$ Institut f\"ur Theoretische Festk\"orperphysik,
 	Universit\"at Karlsruhe, Wolfgang-Gaede-Str. 1, D-76128 Karlsruhe, Germany}
\address{$^3$ Institute for Theoretical Physics, University of Amsterdam,
	Valckenierstraat 65, 1018 XE Amsterdam, The Netherlands}
\address{$^4$ International School for Advanced Studies (SISSA)
	via  Beirut 2-4,  I-34014 Trieste, Italy}
\ead{dechiara@sns.it}

\begin{abstract}
By means of the time-dependent density matrix renormalization group algorithm we study the 
zero-temperature dynamics 
of the Von Neumann entropy of a block of spins in a Heisenberg chain after a sudden 
quench in the anisotropy parameter. In the absence of any disorder the block entropy 
increases linearly with time 
and then saturates. We analyze the velocity of propagation of the entanglement as a 
function of the initial and final anisotropies and compare, wherever possible, our results 
with those obtained by means of Conformal Field Theory. In the disordered case we 
find a slower (logarithmic) evolution which may signals the onset of entanglement 
localization.  
\end{abstract}


\maketitle

\section{Introduction}

The recent interest in aspects common to quantum information and condensed 
matter~\cite{preskill00} has prompted a florishing activity at the border 
of these disciplines, quite distinct until few years ago. It is hard to 
list all the numerous problems addressed so far. Here we only mention the study of 
entanglement in quantum critical systems (see for example~\cite{osterloh02,osborne02,vidal02,
verstraete04,jk-04,vidal04,chen04,calabrese04,lambert04,korepin04,wu04,ijk-04,somma04,dur05,roscilde05,eisert05,
keating05,peschel05,anfossi05,guhne05,wei05, hartmann} and references therein), and the recent advances relating to 
the density matrix renormalization group (DMRG)~\cite{vidal,daley04,white04}, as they are 
tightly connected to the present work. In fact in this paper we apply the recently developed time 
dependent DMRG (t-DMRG)~\cite{vidal,daley04,white04} to the study of the dynamics 
of entanglement entropy in Heisenberg chains.

Among all the various ways to quantify entanglement here we consider the block entropy which has 
recently analyzed in several different situations (see e.g.~\cite{vidal02,calabrese04,korepin04,
eisert05,keating05,peschel05,plenio05,levine04,zhou05,latorre05}).
It has been demonstrated~\cite{hlw94,vidal02,calabrese04} that the entropy of a block of 
spins $S_\ell$ in the ground state of a spin chains is very sensitive to its critical 
properties. In the case of a block with one boundary with the rest of the chain
$S_\ell$, for $\ell\gg1$, diverges at the critical point as $ S_\ell=(c/6) \log_2 \ell $
where $c$ is the central charge of the corresponding conformal field theory (CFT) of the 
model considered. 
In contrast in non critical systems the entropy saturates to the finite value 
$S_\ell=(c/6) \log_2 \xi$, with $\xi\gg1$ the correlation length \cite{calabrese04}. 
In the presence of quenched disorder the properties of the block entropy for critical chains remain 
remarkably universal. 
For the models analyzed in Ref. \cite{refael04} using real-space renormalization group,
it still diverges logarithmically with a ``renormalized'' central charge
$c_{\rm eff}=c\ln 2$~\cite{refael04}, where $c$ is the central charge of the corresponding pure model.
It has been conjectured that such an effective central charge characterizes 
generically the critical systems
with quenched disorder \cite{refael04,laflorencie05}.

The interest in the properties of entanglement in condensed matter has also extended 
to understand its dynamical behaviour.
Like for the case of propagation of excitations in condensed media, it 
recently became of interest to know how entanglement could propagate through spin 
chains. This question was studied by looking at two-particle~\cite{montangero03,amico04,koniorczyk05,sen05} 
and many-particle entanglement~\cite{calabrese05,montangero05}. The dynamics of entanglement was studied 
either by preparing the system in a state (not an eigenstate) with all the entanglement 
localized in a given part of the chain or after a sudden quench of some of the couplings 
of the model Hamiltonian. 

In this paper we consider static and dynamical properties of the entropy of a block 
of spins for an anisotropic Heisenberg chain both in the clean and in the disordered case.
We investigate the problem by means of DMRG~\cite{schoelwock05} for the static part \cite{sleijpen} and 
its time-dependent version~\cite{daley04,white04} for the dynamical evolution. 
The static entropy of the Heisenberg model was considered numerically in Ref.~\cite{vidal02,peschel05,zhao05,zhou05,laflorencie05-bis} 
and some exact results are known in the isotropic ferromagnetic limit~\cite{salerno04} and in the XX and Ising limits \cite{jk-04,ijk-04,peschel05,keating05}.
In this work we consider again this case for completeness. In addition it serves as an important check for the 
other cases considered (see Sections \ref{staticclean}, \ref{dynamicclean}, \ref{staticdisordered},
\ref{dynamicdisordered}) for which these are, to the best of our knowledge, the first numerical 
calculations available for the Heisenberg model. 
The results of the dynamics of an ordered chains are compared with the CFT results of Ref.~\cite{calabrese05}.  In the static disordered case we confirm the prediction of~\cite{refael04}. 
Section \ref{dynamicdisordered} is devoted to the dynamics of a disordered Heisenberg chain. Here we 
predict a slow evolution for the dynamics of entanglement which hints at some sort of entanglement localization. 

\section{The model}

The model that we consider is a  spin chain of length $N$ with {\it open} boundary conditions described 
by the Heisenberg Hamiltonian~\cite{lieb}
\begin{equation} 
\label{eq:ham}
	H=\sum_{i=1}^{N-1} J_i (\sigma_x^i\sigma_x^{i+1}+
	\sigma_y^i\sigma_y^{i+1}+\Delta\sigma_z^i\sigma_z^{i+1})
\end{equation}
where $\sigma_{x,y,z}^i$ are the Pauli matrix operators relative to the $i$th spin; 
$J_i$ are the coupling constants that we assume time independent but possibly 
space-dependent in the random case; finally $\Delta$ is the anisotropy parameter.
The homogeneous chain is critical for $-1\leq\Delta\leq 1$. The corresponding Conformal Field Theory 
(CFT) is described using a central charge $c=1$.

We study the properties of the Von Neumann entropy $S_\ell$ of a block containing 
the first $\ell$ spins. The block is described by the reduced density matrix 
$\rho_\ell={\rm Tr}_{i>\ell} \rho$ and $S_\ell$ is defined as:
\begin{equation}
	S_\ell=-\textrm{Tr}_\ell \left(\rho_\ell \log_2 \rho_\ell\right)
\end{equation}
We used the t-DMRG algorithm with a second order Trotter expansion of $H$ as described in 
Refs. \cite{daley04,white04}. We checked the precision of the numerics by comparing 
the results with the case $\Delta =0$ where Eq. (\ref{eq:ham}) is mapped onto a free fermion 
model~\cite{lieb}.

We first present the result for the homogeneous case and then that for the disordered one. 

\section{Homogeneous chain}

\subsection{Ground state properties}   
\label{staticclean}
\begin{figure}[ht]
\begin{center}
\includegraphics[scale=0.68]{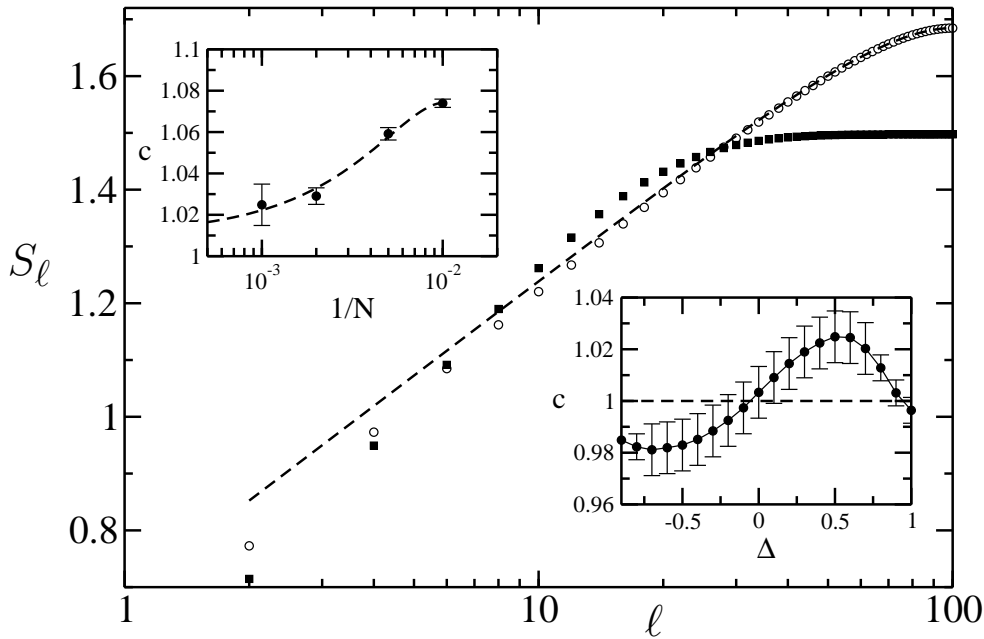}
\caption{The Block entropy $S_\ell$ for $N=200$  for a critical value $\Delta=0.0$ 
	(circles) and non-critical value $\Delta=1.8$ (squares) and $m=120$. The critical 
	data compared with the CFT prediction Eq. (\ref{eq:CFT-static}) (dashed line). 
	Lower inset: central charge extrapolated by fitting the numerical data 
	$S_\ell$ with Eq. (\ref{eq:CFT-static}) for different values of $\Delta$. 
	The data are for $N=1000$ and $m=120$. Upper inset: scaling of $c$ extrapolated as 
	a function of $1/N$ for the worst case $\Delta=0.5$ and compared to a quadratic fit 
	(dashed line).}
\label{fig:1}
\end{center}
\end{figure}

The CFT prediction for a block of length $\ell$ in an open chain of total length $N$ 
is~\cite{calabrese04}
\begin{equation} 
\label{eq:CFT-static}
	S_\ell=\frac{c}{6} \log_2 \left[\frac{N}{\pi} 
	\sin\left(\frac{\pi}{N} \ell\right)\right] +A
\end{equation}
where $A$ is a non-universal constant related to the analogous one in the 
system with periodic boundary condition \cite{calabrese04,zhou05}.  
As already mentioned the static properties of the Heisenberg  
model was considered numerically in Ref.~\cite{vidal02,peschel05,zhao05,zhou05} and here for completeness 
we present the result of our DMRG calculations. They are shown in Fig. \ref{fig:1}. 
The main plot shows the difference in the entropy for a critical and a non critical value of
$\Delta$ and the comparison  to the CFT prediction. 
For large $\ell$ (i.e. $\ell> 10$) where Eq. (\ref{eq:CFT-static}) is expected to work, 
the agreement is very good.
In the lower inset we show our results for the central charge, obtained through  
a fit of numerical data with Eq. (\ref{eq:CFT-static}), for different values of $\Delta$ in the 
critical interval. The fitted $c$ shows small non-universal variations with $\Delta$ which 
decrease increasing $N$ and are expected to vanish in the thermodynamic limit. 
This is shown in the top inset for the worst case $\Delta=0.5$. 
The extrapolated value of the central charge (upper inset) has been obtained fitting $c(1/N)$ with 
a quadratic polynomial and taking the limit $1/N \to 0$. The result is $c=1.01\pm 0.05$.
Note that since we are using open boundary conditions the entropy oscillates with 
the parity of the block. This can be explained with a simple argument. Because of the 
XXZ interaction, neighbor spins tend to form spin singlets, so the oscillations in 
the entropy of block reflect the breaking or not of one of these singlets. Such an alternating behaviour has been observed also in Ref.~\cite{laflorencie05-bis}.
Though this  behavior disappears increasing $N$ we fitted numerical data for even $\ell$ 
because the convergence is faster.

\subsection{Dynamical behaviour}   
\label{dynamicclean}
To date the only results obtained for the evolution of block entropy after a quench have been 
obtained in Ref. \cite{calabrese05}. By means of CFT 
it was shown that a quench of the system from a non critical to a critical point leads the block  
entropy to increase in time until a saturation point is obtained. 
For periodic boundary conditions, the time at which the entropy saturates is given by $t^*=\ell/(2v)$ 
where $v$ is the spin wave velocity: $v=\partial E_k/\partial k |_{k=0}$.
This phenomenon has a simple interpretation in terms of quasiparticles excitations emitted from the 
initial state at $t=0$ and freely propagating with velocity $v$.
This explanation holds even for non critical systems, as it has been confirmed by the exact solution of
the Ising chain dynamics \cite{calabrese05}. 
However, since in lattice models there are particles moving slower than $v$, after $t^*$ the entropy 
does not saturate abruptly, but is a slowly increasing function of the time.
We now consider the dynamics of the Heisenberg model with open boundary conditions, and we will find
that $t^*=\ell/v$. At the end of this section we will interpret this result in terms of quasiparticles
and we will derive it within CFT.
 
\begin{figure}[ht]
\begin{center}
\includegraphics[scale=0.5]{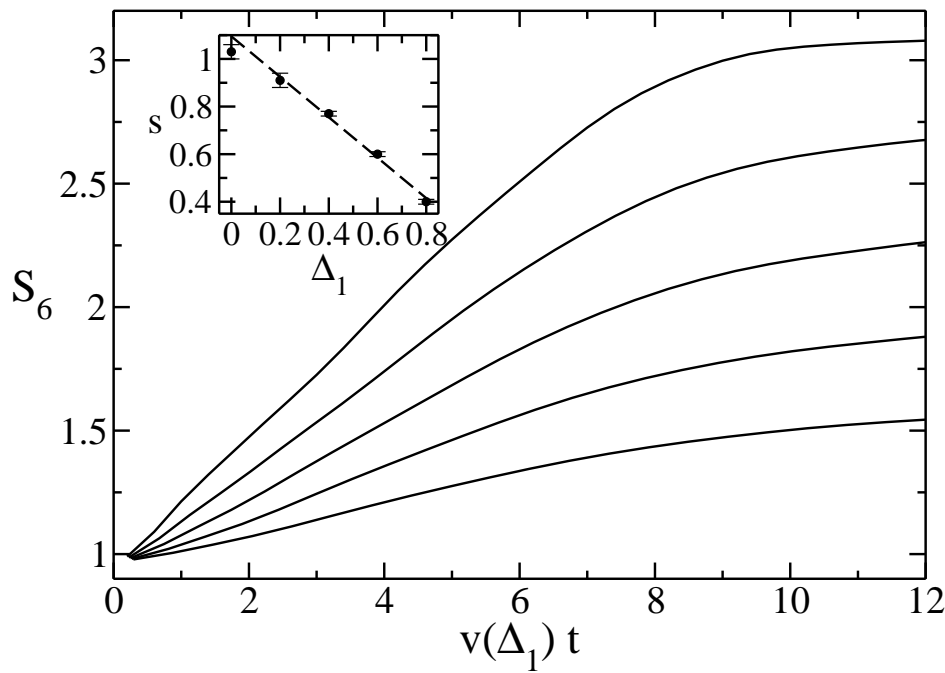}
\caption{Evolution of the entropy $S_{6}$ with various quenches. $\Delta_0=1.5$ while 
	$\Delta_1=0.0, 0.2, 0.4, 0.6, 0.8$ as a function of
        $v(\Delta_1) t$. Inset: initial slope
	value of $S_{6}$ as a function of $\Delta_1$ and comparison to a linear fit with 
	slope $-0.85\pm0.02$ (dashed line).}
\label{fig:2}
\end{center}
\end{figure}

In the case of the Heisenberg Hamiltonian in the critical regime 
the spin wave phase velocity is given by
$ v(\Delta)=2 J\pi\sin\theta /\theta $ with $ \cos\theta=\Delta $~\cite{korepin}.
The initial state of the system is chosen as the ground state of Hamiltonian Eq. (\ref{eq:ham}) 
with $\Delta=\Delta_0>1$, after a quench the system evolves with same Hamiltonian but with a 
different anisotropy $\Delta=\Delta_1 \in [0;1]$.
In the simulations we considered chains with  $N=50$, a Trotter slicing $J \delta t=5\cdot 10^{-2}$ and a 
truncated Hilbert space of $m=200$. The block was chosen to be of 6 sites which is large enough, 
as we show, to confirm the CFT prediction.
We have checked convergence with $m$ and $\delta t$. For the special case 
$\Delta_0=+\infty$ and $\Delta_1=0$ we compare our data with the exact result obtained diagonalizing the 
$XX$ model as shown in  \ref{appen}.

\begin{figure}[ht]
\begin{center}
\includegraphics[scale=0.5]{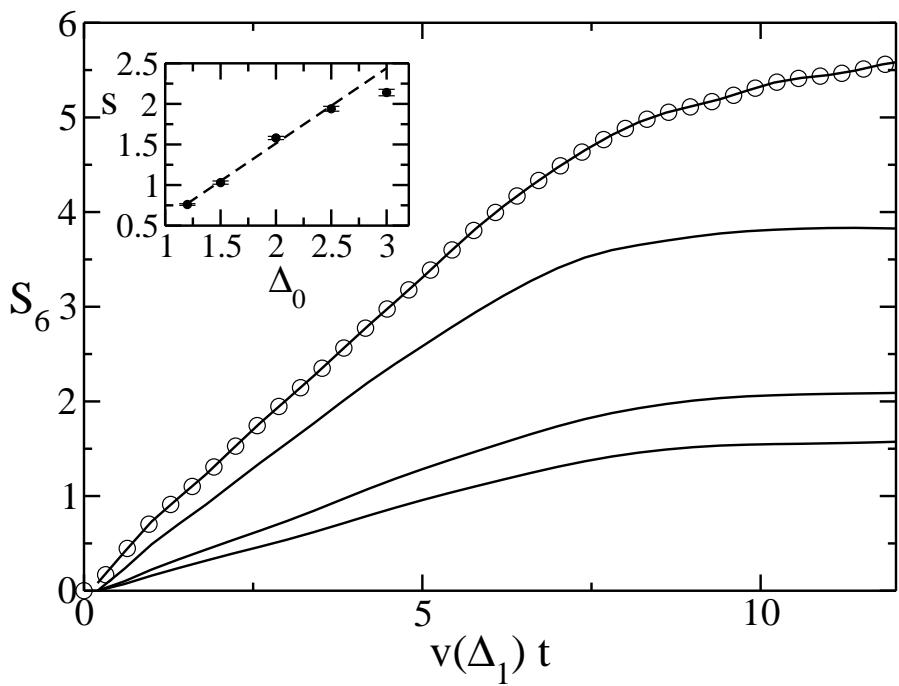}
\caption{Evolution of the entropy $S_{6}$ with various quenches. $\Delta_0=1.2,1.5,3.0, \infty$ 
	while $\Delta_1=0.0$ as a function of  $v(\Delta_1)t$ and shifted so to coincide in $t=0$. 
	For $\Delta_0=\infty$ we show also the exact result obtained by diagonalization (circles ).
	Inset: initial slope value of $S_{6}$ as a function of $\Delta_1$.}
\label{fig:3}
\end{center}
\end{figure}

\begin{figure}[ht]
\begin{center}
\includegraphics[scale=0.5]{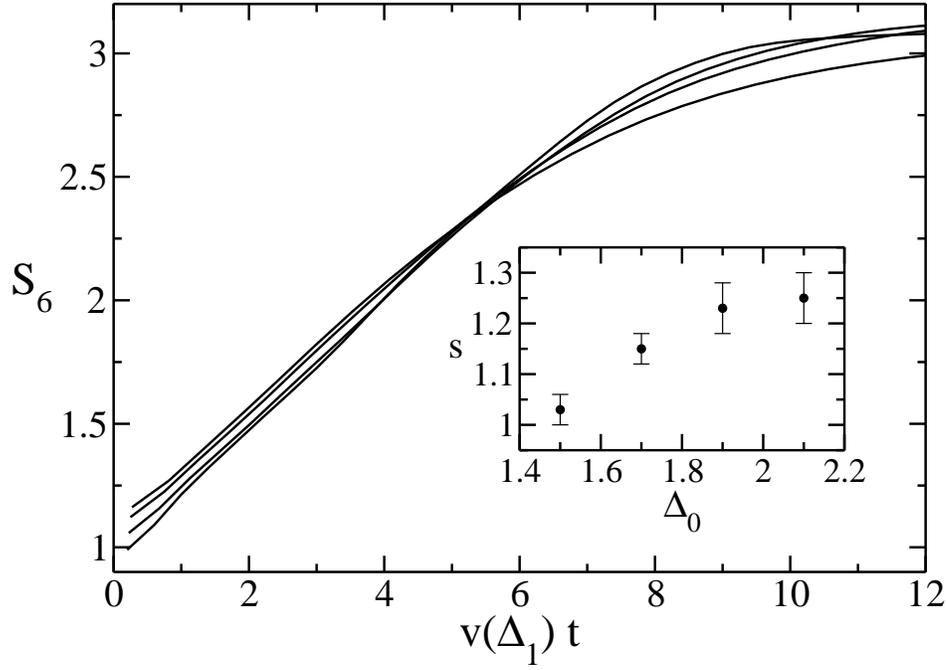}
\caption{Evolution of the entropy $S_{6}$ with fixed quench as a function 
	of $v(\Delta_1) t$.  Fixed quench $\Delta_0-\Delta_1=1.5$ for $\Delta_0=1.5, 1.7, 1.9,2.1$. 
	Inset: initial slope value of $S_{6}$ as a function of $\Delta_0$.}
\label{fig:4}
\end{center}
\end{figure}

Various quenches have been considered. First of all we present the results with $\Delta_0$ fixed 
and $\Delta_1$ variable shown in Fig.~\ref{fig:2}. As in
Ref. \cite{calabrese05} the entropy grows linearly in time $S \sim s Jt$ and finally
saturates. 
At $t=t^*=\ell/v$, the entropy does not saturate abruptly. 
As discussed for the Ising model, this is due to slow quasiparticles \cite{calabrese05}.
The inset of Fig.~\ref{fig:2} shows the initial slope as a function of 
$\Delta_1$. The dashed line is a linear fit with slope $0.85$ on the
data except $\Delta_1=0$. 
In Fig. \ref{fig:3} the time evolution of the entropy is shown for
quenches starting from different values of $\Delta_0$ and ending with the same
$\Delta_1=0$. 
The case in which both $\Delta _0$  and  $\Delta _1$ are varied (keeping their difference 
fixed, $\Delta_0-\Delta_1=1.5$) is shown in Fig.~\ref{fig:4}.  

\begin{figure}[ht]
\begin{center}
\includegraphics[scale=1]{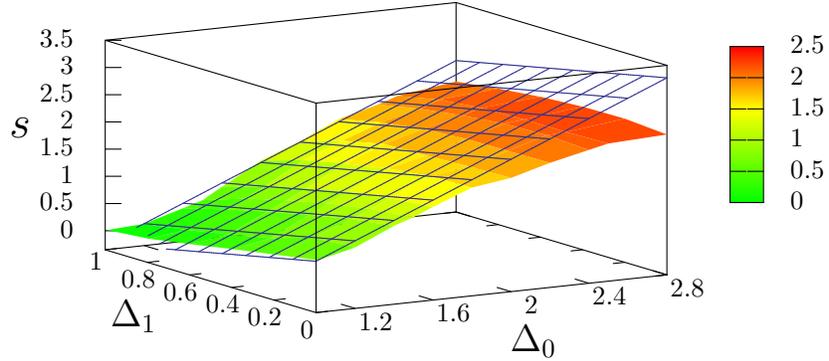}
\vspace{-1.5cm}
\caption{Derivative $s$ of the entropy with respect to time as a
  	function of $\Delta_0$ and $\Delta_1$ (surface) and compared to a linear fit (grid lines):
  	$s = 1.50 \Delta_0 -0.84 \Delta _1 -0.90$. The deviation with
  	the linear fit is less than $10\%$  for $(\Delta_0,\Delta_1)\in[1.4,2.4]\times[0.4,0.8]$.} 
\label{fig:5}
\end{center}
\end{figure}

The behaviour of the slope $s$ (the entropy for $t \le t^*$ goes as $S \sim s Jt$ ) for $\ell=20$
as a function of $\Delta _0$ and $\Delta_1$ can be fitted for a large portion of values by the law 
$$
s = 1.50 \Delta_0 -0.84 \Delta _1 -0.90 \,\,\,\, .
$$
In Fig. \ref{fig:5} the fit and the numerical data are compared.  
Following the simple model introduced in 
\cite{calabrese05}, the initial slope of the entropy increase is a
non-trivial function of both $\Delta_0$ and $\Delta_1$ as it depends
both on the crossection for producing the quasiparticles and their velocity. 

According to the argument reported in Ref. \cite{calabrese05}, the block entropy is proportional to 
the numbers of pairs of entangled quasiparticles, emitted from any point at the time $t=0$, of which one reaches  
the block and the other reaches the rest of the system. If these particles are all moving at speed $v$, the 
entropy of a block in a chain infinite in both directions is at first linear in $t$ until $t=t^*=l/(2v)$ and 
then saturates to a value proportional to $\ell$. 
Clearly the presence of boundary condition at $x=0$, changes this result, since it acts as a wall and 
particles reaching it cannot propagate any longer in a straight line. 
If we assume that the wall at $x=0$ is reflecting, it works like a mirror for the quasiparticles. 
The resulting ``effective length'' the block is $2\ell$ and the saturation time $t^*=2\ell/(2v)$, in 
agreement with what we have found numerically. 
On the other hand a completely adsorbing wall does not change the ``effective length'' the block and 
the resulting saturation time is the same as for periodic conditions.

It is very simple to generalize the CFT calculation of Ref. \cite{calabrese05} to the case of a 
boundary condition that is of the same kind of the initial condition (e.g. fixing the spins in the 
same direction both at initial time and at $x=0$, or having a disordered initial state and free 
boundary condition at $x=0$). 
In this case, as shown in~\cite{calabrese04}, ${\rm Tr}\rho_\ell$ transforms like a one-point function.
The imaginary-time half-strip of width $2\epsilon$ (in the notation of Ref. \cite{calabrese05} to which 
we refer the reader for details) is mapped on the half-plane by the conformal transformation 
$z(w)=\sin(\pi w/(2\epsilon))$ (more details about this calculation can be
found in Ref.~\cite{calabrese06} where it has been performed in a different
context). Continuing the one-point function obtained to real time
and performing the ``replica trick'' \cite{calabrese04}, we easily obtain $t^*=\ell$, in agreement with 
the quasiparticles interpretation and our numerical results.

\section{Disordered chain}

\subsection{Ground state properties}
\label{staticdisordered}
\begin{figure}[ht]
\begin{center}
\includegraphics[scale=0.68]{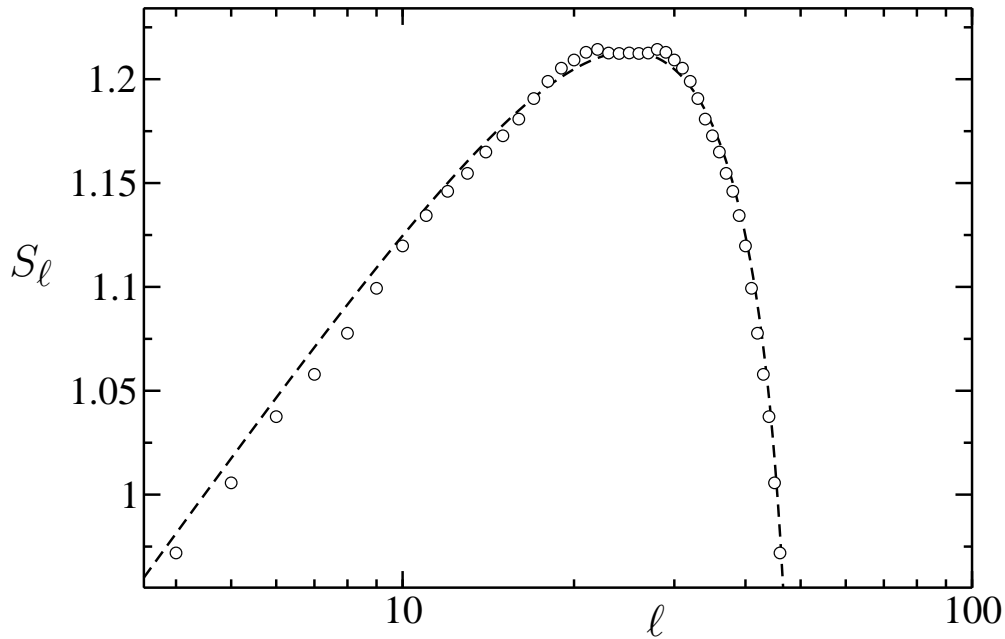}
\caption{The Block entropy $S_\ell$ for the random model  for a critical 
	value $\Delta=1.0$ (circles) for $N=50$ and $m=50$.  These are compared 
	with the Refael and Moore prediction Eq. (\ref{eq:CFT-static}) (dashed line). The data 
	have been averaged over $10^4$ realizations.}
\label{fig:7}
\end{center}
\end{figure}

The striking prediction of Refael and Moore~\cite{refael04} that the block entropy of random chains 
has a logarithmic divergence with an effective (universal) central charge 
has been confirmed by numerical calculations on the $XX$ model~\cite{laflorencie05}.
Here we extend these results to the model defined in Eq. (\ref{eq:ham}) where the 
randomness is introduced through the couplings $J_i$ chosen to be random numbers 
with a uniform distribution in the interval $[0;1]J$.  
According to Ref.~\cite{refael04}, the block entropy  should scale with an effective central charge 
$c_{\rm eff}=\ln 2\simeq 0.69$.
In Fig. \ref{fig:7} we show for $N=50$ the entropy of a block of
length $\ell$ for the random model for $\Delta=1$ averaged over $10^4$ different
configurations of disorder. 
From the scaling of the entropy as a function of $\ell$ we can extract the 
effective central charge for this model. The result is $c_{\rm fit} = 0.67 \pm
0.05$ in very good agreement with Refael and Moore proposal.

\subsection{Dynamical behaviour}
\label{dynamicdisordered}

No analytic results are known so far for the dynamics of block entropy in the 
case of disordered systems. This, in our opinion may be an interesting question as it 
is well known that in the presence of disorder the ballistic propagation of quasiparticles 
turns into diffusion and, in certain circumstances, into localization.
%
\begin{figure}[ht]
\begin{center}
\includegraphics[scale=0.5]{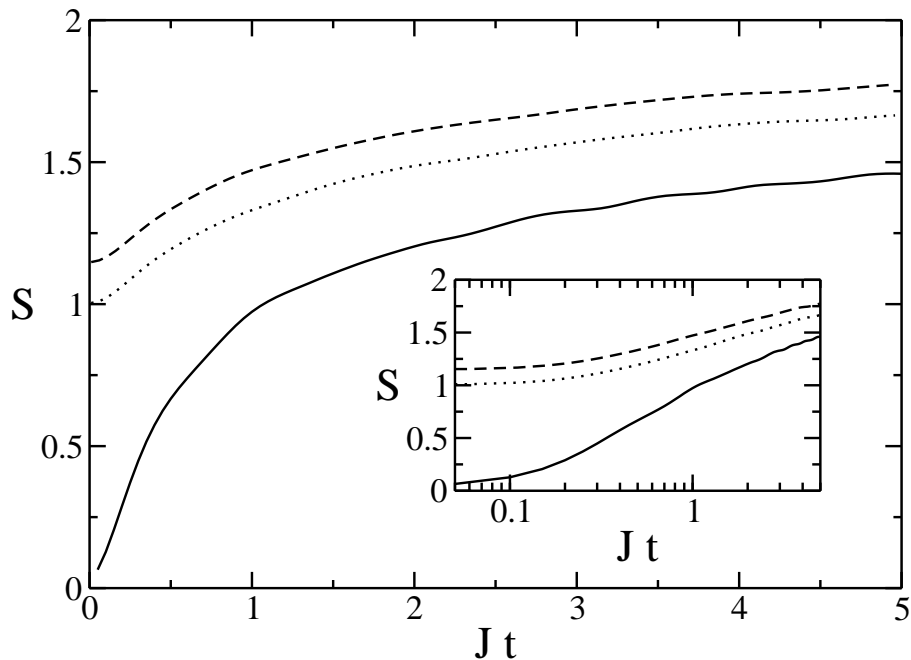}
\caption{Dynamics of the entropy for the random Heisenberg model obtained using the 
	t-DMRG  for various quenches. The solid line is a quench from $\Delta_0=\infty$ 
	to $\Delta_1=0$ for a block of $\ell=10$. The dotted and 
	dashed lines are quenches from $\Delta_0=2$ to $\Delta_1=0$ for two block sizes 
	$\ell=10$, $\ell=20$ respectively. In the three cases we considered $N=50$.  
	Inset: the same plot but in semi-logarithmic scale. 
	The parameters of the DMRG calculation are $m=60$ and 
	$J \delta t=5\cdot10^{-2}$. 
	The data have been averaged over $10^3$ for the quench from  $\Delta_0=+\infty$ and 
	$400$ for those with $\Delta_0=2$.}
\label{fig:8a}
\end{center}
\end{figure}
%
It is therefore to be expected that entanglement itself will be affected by the presence 
of static randomness.  At the level of the ground state properties the effect of disorder 
manifests in a ``renormalization" of the central charge. As we will show below the dynamical 
behaviour, instead, is strikingly different in the clean and disordered cases.  

As in the clean case we analyzed the evolution of a random chain with a quench in 
the anisotropy from a non-critical $\Delta_0$ to a critical value $\Delta_1$ for 
various cases.  Interestingly now the entropy does not grow linearly 
as in the non-random case. Although it is very difficult, in the absence of any analytic 
result, to ascertain the exact  time dependence of the entropy, our data clearly  indicate that 
the entropy grows logarithmically as a function of time.
This is  shown in Fig. \ref{fig:8a} where we report the t-DMRG results for 
several quenches. After a transient behaviour, all the curves in 
Fig.~\ref{fig:8a} behave like
\begin{equation}
S_\ell\sim \kappa \ln Jt\,,
\end{equation}
where $\kappa$ depends on the details of the quench. For example we 
find $\kappa\sim0.5$ for $\Delta_0=\infty$ and $\Delta_1=0$, and 
$\kappa\sim0.22$ for $\Delta_0=2$ and $\Delta_1=0$ (with $\ln$ the natural
logarithm). Clearly such logarithmic growth cannot continue indefinitely,
but with t-DMRG it is hard to investigate times larger than those reported.

\begin{figure}[ht]
\begin{center}
\includegraphics[scale=0.5]{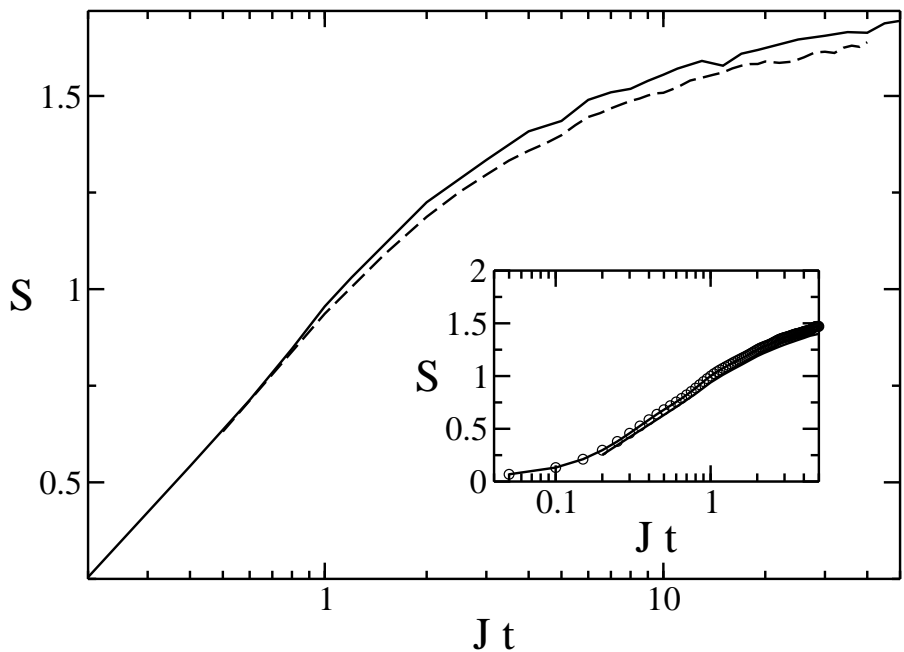}
\caption{Dynamics of the entropy for the random Heisenberg model.  
	Two quenches from $\Delta_0=+\infty$ to $\Delta_1=0$ for different lengths of the block:
	$\ell=30$ (dashed line) and $\ell=50$ (solid line). In both cases the length of the chain is 
	$N=120$ and the number of configurations for the averaging was $10^4$.   
	The curves presented in this figure were obtained by the diagonalization of the $XX$ model.
	A comparison between the t-DMRG and the exact diagonalization is presented in the inset for 
	the case $\ell=10$ and $N=50$.}
\label{fig:8b}
\end{center}
\end{figure}

To shed some light on the long time behaviour of $S_\ell$, we consider
extensively the diagonalization of the XX model starting from 
$\Delta_0=\infty$, as described in the appendix.
The exact diagonalization of the $XX$ model allows us to consider chain up to 
$120$ sites, blocks of up to $50$ sites and to follow the dynamical evolution 
to longer times as presented in Fig. \ref{fig:8b}.
The analysis on the $XX$ model was used furthermore to check the accuracy of 
the t-DMRG data. 
This is shown in the inset of Fig. \ref{fig:8b} where the discrepancy is below $3\%$.

\begin{figure}[ht]
\begin{center}
\includegraphics[scale=0.68]{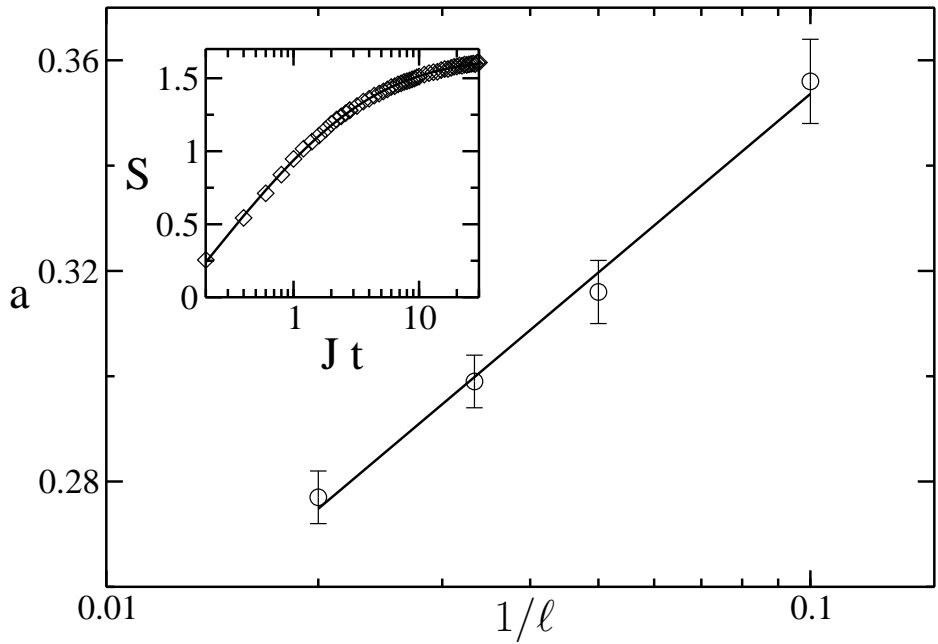}
\caption{Scaling of the coefficient $a(\ell)$ of Eq.(\ref{logtime}) with the dependence of 
	the length of the block. A power law behaviour $a(\ell)\sim \ell^{-\nu}$ (with  $\nu=0.16\pm 0.01$)
	fits well all the range considered. In the inset we show an example of a fit with the 
	logarithmic behaviour suggested in Eq. (\ref{logtime}); the accuracy shown in the figure is 
	obtained for all the cases analyzed in this work}
\label{fig:9}
\end{center}
\end{figure}

Interestingly, Fig. \ref{fig:8b} shows the end of the logarithmic growth of 
$S_\ell$, even if the complete saturation is not yet reached.
All the data of Fig. \ref{fig:8b} are well fitted by the function
\begin{equation} \label{log-fit}
	S^{(\infty,0)}_\ell = -\frac{1}{2} \ln\left (\frac{1}{t}+a(\ell)\right)+b(\ell)
\label{logtime}
\end{equation}
An example of this fit is given in the inset of Fig. \ref{fig:9}.
We find that the parameter $b(\ell)$ is slightly depending on $\ell$, being
approximately $1.1$ for all the cases considered.
Instead $a(\ell)$ is a decreasing functions of $\ell$, as shown in 
Fig. \ref{fig:9} for $\ell=10,20,30,50$. We note that our data can be fitted
by the simple scaling behaviour $a(\ell) \sim \ell ^{-\nu}$ with 
$\nu=0.16\pm 0.01$, as shown in Fig. \ref{fig:9}. 
At this stage however a logarithmic dependence on $\ell$ cannot be excluded. 
However, assuming that Eq. (\ref{log-fit}) is true for very large times, 
and assuming that the power law behaviour of $a(\ell)$ is corrected, we have 
that for $t\to\infty$ the entanglement entropy saturates to 
$\nu/2 \ln\ell$. Thus for infinitely large times the entanglement entropy 
saturates to a value that is reminiscent of the one in the ground state,
but the prefactor of the logarithm seems to be different (if our fit 
holds for large times it is exactly the half).

Several considerations are in order at this stage. 
As compared to the clean case (Figs. \ref{fig:2}, \ref{fig:3}, \ref{fig:4}) 
the increase of the entropy as a function of time is much slower. 
The logarithmic behaviour does not follow from 
an extension of the argument for the clean case~\cite{calabrese05} assuming 
that pair of particles that are emitted diffuse rather than moving 
ballistically. At this stage we cannot present
arguments to derived the logarithmic increase of entropy in disordered chain. 
This behaviour is probably associated with a sort of entanglement localization.

%

\section{Conclusions}
In this paper we have presented a detailed account of the static and the dynamics 
of the block entropy for the Heisenberg model. For the static case
we confirm the results of analytic calculations, i.e. the entropy diverges 
logarithmically and that the prefactor is proportional to the central charge 
$c=1$ in the clean case and $c_{\rm eff}=\ln 2$ in the disordered case.
For the dynamical case we have confirmed the CFT prediction that the entropy should 
grow linearly in time and that the saturation time $t^*=\ell/v$ as a consequence of 
a reflecting wall at the boundary of the chain.
The dynamical behaviour of the entropy after a quench depends {\em qualitatively}
on the presence of disorder. Unlike for the clean case in the disordered case 
the entropy grows logarithmically with time signaling the possibility of  entanglement 
localization.

\ack
We want to thank M. Rizzi, and D. Rossini for very useful discussions.
This work was supported by IBM (2005 Faculty award), and by the European Community 
through grants RTNNANO, SPINTRONICS and SQUBIT2. The present work has been 
performed within the ``Quantum Information" research program of Centro di Ricerca 
Matematica ``Ennio De Giorgi'' of Scuola Normale Superiore.
PC acknowledges support from the Stichting voor Fundamenteel Onderzoek der Materie (FOM). SM ackonwledges supports from Alexander Von Humboldt Foundantion.

\appendix
\section{}
\label{appen}

In order to check our numerical results we compared them with the available 
analytic solution of the $XX$ model where it reduces to a free fermion system. For 
completeness in this appendix we sketch how to calculate the entropy for the $XX$ model with \emph{open} 
boundary conditions. For $\Delta=0$ the Hamiltonian Eq. (\ref{eq:ham}) can be rewritten, using 
the Jordan-Wigner transformation \cite{lieb}, as:
\begin{equation} \label{eq:ham1}
H=2\sum_{j=1}^{N-1}J_j\left( c_j^\dagger c_{j+1} + h.c.   \right)
\end{equation}
where $c_j^\dagger=(\sigma^j_x-i\sigma^j_y)$.
It is possible to rewrite Hamiltonian Eq. (\ref{eq:ham1}) in diagonal form
\begin{equation} \label{eq:ham2}
H=\sum_{k=1}^{N} E_k b_k^\dagger b_k
\end{equation}
where the new ladder operators are connected to the old ones with an orthonormal transformation:
\begin{equation}
b_k=\sum_{j=1}^N a_{kj}c_j
\end{equation}
The matrix $\{a_{kj}\}$ contains the normalized eigenvectors of the $N\times N$ adjacency matrix
$\{J_{kj}\}$ whose elements are defined as $J_{1j}=2J_1 \delta_{j,2}$, $J_{Nj}=2J_{N-1} \delta_{j,N-1}$
$J_{kj}=2J_k \delta _{k,j+1}+ 2 J_{k-1}\delta _{k,j-1}$ for $k=2,\dots ,N-1$.

To calculate the entropy of a block of size $\ell$ we use  \cite{laflorencie05}:
\begin{equation}
\label{entro}
	S=-\sum_{\alpha=1}^\ell \lambda_\alpha\log_2\lambda_\alpha
	+ (1-\lambda_\alpha)\log_2(1-\lambda_\alpha)
\end{equation}
where $\lambda_\alpha$ are the eigenvalues of the correlation matrix 
$C_{ij}=\langle c^\dagger_i c_j\rangle,\quad i,j=1,\ell$.

As an example we consider the time dependence of the block entropy after a quench 
from $\Delta_0=\infty$ to $\Delta_1=0$. The initial state is the antiferromagnetic state:
\begin{equation}
	\ket{\psi_0}=c^\dagger_1 c^\dagger_3 \cdots c^\dagger_{N-1}\ket 0
\end{equation}
The correlations are easily evaluated using the transformation $\{a_{kj}\}$ with the result
\begin{eqnarray}
\label{corr}
	\langle c^\dagger_i(t) c_j(t)\rangle&=& \sum_{k,k'} a_{ik}a_{jk'}e^{-i(E_k-E_k')t} 
	\bra{\psi_0}b^\dagger_k b_{k'}\ket{\psi_0}=\nonumber\\
	&=& \sum_{k,k'}\sum_{i'j'} a_{ik}a_{jk'}a^*_{i'k}a^*_{j'k'}e^{-i(E_k-E_k')t} 
	\bra{\psi_0}c^\dagger_{i'} c_{j'}\ket{\psi_0}=\nonumber\\
	&=& \sum_{k,k'}\sum_{i'=1,3}^{N-1} a_{ik}a_{jk'}a^*_{i'k}a^*_{i'k'}e^{-i(E_k-E_k')t}
\end{eqnarray}
By using Eq.(\ref{corr}) to calculate the eigenvalues of the correlation matrix, the block 
entropy in readily evaluated from Eq.(\ref{entro}).
For example when the chain is homogeneous $J_i=J$ we obtain:
\begin{equation}
a_{kj}=\sqrt{\frac{2}{N+1}}\sum_{j=1}^N \sin kj
\end{equation}
and the corresponding energy levels are $E_k=4 J\cos k_n$ where $k_n=(\pi n)/(N+1), \quad n=1,2,\dots,N$.


\end{document}